\documentclass[10pt,conference,letter]{IEEEtran}

\usepackage{graphicx}
\usepackage{epstopdf}
\usepackage{float,url}
\usepackage[caption=false,font=footnotesize]{subfig}
\usepackage{amsmath,epsfig,psfrag,amssymb} %subfigure
\usepackage{amsfonts,graphics,graphicx}
\usepackage{algorithm}
\usepackage{algorithmicx,algpseudocode} 
\usepackage{textcomp}
\usepackage{xcolor}
\usepackage{booktabs}
\usepackage{balance,color}

\newtheorem{insight}{\textbf{Insight}}
\newcommand{\ignore}[1]{}

\usepackage{tabularx}
\usepackage{cite}
\begin{document}
%\tableofcontents

%\title{A Case Study on Detecting COVID-19 Themed Malicious Websites}
\title{Data-Driven Characterization and Detection of COVID-19 Themed Malicious Websites}

%\tableofcontents
%\title{Countering COVID-19 Themed Malicious Websites: A Case Study}

\ignore{
\author{\IEEEauthorblockN{Mir Mehedi Ahsan Pritom}
\IEEEauthorblockA{Department of Computer Science\\
University of Texas at San Antonio\\
Email: mirmehedi.pritom@utsa.edu}
\and
\IEEEauthorblockN{Shouhuai Xu}
\IEEEauthorblockA{Department of Computer Science\\
University of Texas at San Antonio\\
Email: shouhuai.xu@utsa.edu}
\and 
\IEEEauthorblockN{Min Xu}
\IEEEauthorblockA{Mastercard\\
Email:min.xu@mastercard.com}}
}

\author{\IEEEauthorblockN{Mir Mehedi Ahsan Pritom\IEEEauthorrefmark{1},
Kristin M. Schweitzer\IEEEauthorrefmark{2},
Raymond M. Bateman\IEEEauthorrefmark{2},
Min Xu\IEEEauthorrefmark{3} and
Shouhuai Xu\IEEEauthorrefmark{1}}
\IEEEauthorblockA{\IEEEauthorrefmark{1}Department of Computer Science, University of Texas at San Antonio}
\IEEEauthorblockA{\IEEEauthorrefmark{2}U.S. Army Research Laboratory South - Cyber}
\IEEEauthorblockA{\IEEEauthorrefmark{3}Mastercard}
}

%\author{Lars Th{\o}rv{\"a}ld}
%\affiliation{%
%  \institution{The Th{\o}rv{\"a}ld Group}
%  \streetaddress{1 Th{\o}rv{\"a}ld Circle}
%\email{larst@affiliation.org}

\maketitle

\begin{abstract}
COVID-19 has hit hard on the global community, and organizations are working diligently to cope with the new norm of ``work from home''. However, the volume of remote work is unprecedented and creates opportunities for cyber attackers to penetrate home computers. Attackers have been leveraging websites with COVID-19 related names, dubbed {\em COVID-19 themed malicious websites}.  These websites mostly contain false information, fake forms, fraudulent payments, scams, or malicious payloads to steal sensitive information or infect victims' computers. 
In this paper, we present a data-driven study on characterizing and detecting COVID-19 themed malicious websites. Our characterization study shows that attackers are agile and are deceptively crafty in designing geolocation targeted websites, often leveraging popular domain registrars and top-level domains. Our detection study shows that the Random Forest classifier can detect COVID-19 themed malicious websites based on the lexical and WHOIS features defined in this paper, achieving a 98\% accuracy and 2.7\% false-positive rate. 
\end{abstract}


\ignore{
}
 
\section{Introduction}
The COVID-19 pandemic has incurred many new cyber attack vectors. Many of these cyber attacks incorporate COVID-19 themed factors into phishing, malware, and scamming schemes for various malicious goals (e.g., monetary benefits, stealing credentials, stealing credit card numbers, or identity theft). For example, there is reportedly a 148\% increase in ransomware attacks in March 2020 compared with February 2020 \cite{VMware_carbonBlack_report}, where many attacks are initiated by malicious websites abusing victims' trust. 

This paper focuses on one emerging attack vector, namely malicious websites leveraging COVID-19 as a theme or {\em COVID-19 themed malicious websites} \cite{helpnetsecurity_covid_websites}. As organizations incorporate the ``work from home'' policy, the consequences of COVID-19 themed malicious websites can be significantly amplified because home computers are often more vulnerable to attack than work computers.
During the COVID-19 pandemic, many people lost their jobs and are affected by mental health issues, which causes excessive pressures. These pressures may make average users even more vulnerable to social engineering attacks waged via COVID-19 themed malicious websites. This increases the motivation of the importance of understanding and defending against COVID-19 themed malicious websites, which is a new problem that has not been studied before in a systematic way.

\noindent{\bf Our contributions}.
In this paper, make the following contributions. First, we propose a methodology for characterizing and detecting COVID-19 themed malicious websites through a data-driven approach. To the best of our knowledge, this is the first study on {\em data-driven} characterization and detection of COVID-19 themed malicious websites. 
Second, we apply the methodology to specific datasets to draw the following insights: (i) some attackers may be incentivized to use cheaper registrars for registering COVID-19 themed malicious websites; (ii) attackers often abuse popular top-level domains for their COVID-19 themed malicious websites; (iii) attackers are agile in waging the COVID-19 themed malicious website attack; (iv) attackers are crafty in using COVID-19 themed keywords, and geographical information in creating COVID-19 themed malicious website domain names; (v) the small degree of data imbalance does not have any significant impact in the effectiveness of detecting COVID-19 themed malicious websites; and (vi) COVID-19 themed malicious website detectors must consider WHOIS features and Random Forest performs better than $K$-nearest neighbor, decision tree, logistic regression, and support vector machine.

\noindent{\bf Paper outline}. The rest of the paper is organized as follows. 
%Section \ref{sec:related-work} discuss and compare other related works. 
Section \ref{sec:related-work} explores the related work. Section \ref{sec:methodology} explores the research questions which guide us to characterize and detect COVID-19 themed malicious websites. Section \ref{sec:experiments} reports the experiments and results. Section \ref{sec:discuss} discuss our weakness and future research opportunities. 
%general defense strategies. %Section \ref{sec:case-study} presents a case study on detecting COVID-19 themed malicious websites.
%Section \ref{sec:related-work} reviews related prior studies. 
Section \ref{sec:conclusion} concludes the paper.

\section{Related Works}
\label{sec:related-work}

%\subsection{Defence against malicious URLs and websites}
%In literature, we find end-to-end frameworks for automated phishing URL detection in real-time \cite{AutomatedPhishingDetection}. However, in our case, we are dealing only with the domain name, and no path information is provided, which is more challenging because of the possibility of increasing false-positive rates. Moreover, their framework has 87\% real-time detection accuracy only for phishing URLs, while results for other attacks leveraging malicious websites are not mentioned with similar approaches. 
\ignore{
%rewrite this section by only considering the prior studies that deal with X-themed attacks, for which there should be very few references. then, compare their perspectives against ours ... no need to say much about comparison at the means level; rather, we stay at the goals-level comparsion

}

Although the problem of COVID-19 themed malicious websites has not been investigated until now, the problem of malicious websites has been studied in the literature prior to the COVID-19 pandemic. The problem of detecting malicious URLs generated by domain generating algorithms has been investigated in \cite{DLSTM_dga_detection}. The problem of detecting phishing websites has been addressed via various approaches, including: the descriptive features-based model \cite{Christou2020PhishingUD}, the lexical and HTML features-based model \cite{DQLearning_PhishingDetection}, the HTML and URL features-based model \cite{Li2019ASM}, and the natural language processing and word vector features-based model \cite{SAHINGOZ2019345}. The problem of detecting malicious websites has been addressed via the following approaches: leveraging application and network layers information \cite{XuCodaspy13-maliciousURL}, leveraging image recognition \cite{MaliciousWebImage_CNN}, leveraging 
generic URL features \cite{DBLP:journals/tist/MaSSV11,generic_features}, leveraging character-level embedding or keyword-based recurrent neural networks \cite{whats_in_url,farhan_douksieh_abdi_2017_1155304,mal_url_rnn}, the notion of adversarial malicious website detection \cite{cns14MaliciousWebsiteXu}.
However, these studies do not consider
features pertinent to the COVID-19 pandemic, which are we leverage. Nevertheless, the present study fall under the umbrella of cybersecurity data analytics \cite{XuAgility2019,XuIEEETIFS2013,
%XuInTrust2014,
XuTIFS2015,XuPLoSOne2015,
%XuVineCopula2015,XuMarkerPointProcess2017,XuJournalAppliedStatistics2018,
XuTIFSDataBreach2018}, which in turn belong to the Cybersecurity Dynamics framework  \cite{XuBookChapterCD2019,XuIEEETNSE2018,XuHotSoS2018Firewall,Pendleton16,XuHotSoS2018Diversity}.
%,XuHotSOS14-MTD,XuInternetMath2015ACD,XuHotSoS2015}, 
%while leveraging pertinent metrics \cite{8017389,XuAgility2019,XuSTRAM2018ACMCSUR}.

\ignore{
\subsection{Studies Related to COVID-19 Themed Attacks}
There are recent studies on the types of cyberattacks and their attack trends during the COVID-19 pandemic %(e.g., ransomware, malicious apps, malicious websites). 
\cite{ten_deadly_covid19_security_threats,CyberCrime_pandemic2020,collier2020implications_policy}, studies on specific cyberattacks related to COVID-19 (e.g., mobile malware ecosystem \cite{COVID_malicious_app_paper}, %,lallie2020cyber_COVID19Paper}, 
%investigate COVID-19 themed mobile malware ecosystem. 
cybercrimes \cite{influence_model_COVID_cybercrime}, fake news related to COVID-19 \cite{CyberCrime_in_time_of_plauge}, and COVID-19 themed cryptocurrency scams \cite{xia2020dont_cryptocurrency}).
%studies a multi-level influence model of COVID-19 themed cybercrime scams to understand the emotional appeal and influence techniques used by attackers. 
To the best our knowledge, this the first study to address an active defense approach against the COVID-19 themed malicious websites. %Hence, in this paper, we study an active defense approach against the COVID-19 themed malicious websites. 
}

%\subsection{Paper Outline}
%Section \ref{sec:methodology} describes our methodology. Section \ref{sec:experiments} presents our experimental results. Section \ref{sec:conclusion} concludes the paper.

\section{Methodology}
\label{sec:methodology}

Our methodology for {\em data-driven} characterization and detection of COVID-19 themed malicious websites is centered at answering a range of research questions.

\subsection{Characterization Methodology}
\label{sec:rq_characterize}
In order to characterize COVID-19 themed malicious websites, we address 4 Research Questions (RQs):
\begin{itemize}
\item RQ1: Which WHOIS registrars are most abused to launch COVID-19 themed malicious websites?
\item RQ2: Which Top Level Domains (TLDs) are most abused by COVID-19 themed malicious websites?
\item RQ3: What trends are exhibited by COVID-19 themed malicious websites?
\item RQ4: Which theme keywords are mostly abused by attackers, and how?
\end{itemize}
We consider WHOIS information because it has shown to be useful in the era prior to the COVID-19 pandemic \cite{XuCodaspy13-maliciousURL,cns14MaliciousWebsiteXu}. Answering the preceding questions will deepen our understanding of COVID-19 themed malicious website attacks.

\subsection{Detection Methodology}
\label{sec:rq_detection}
\begin{figure}[!t]
\centering
\includegraphics[width=.49\textwidth]{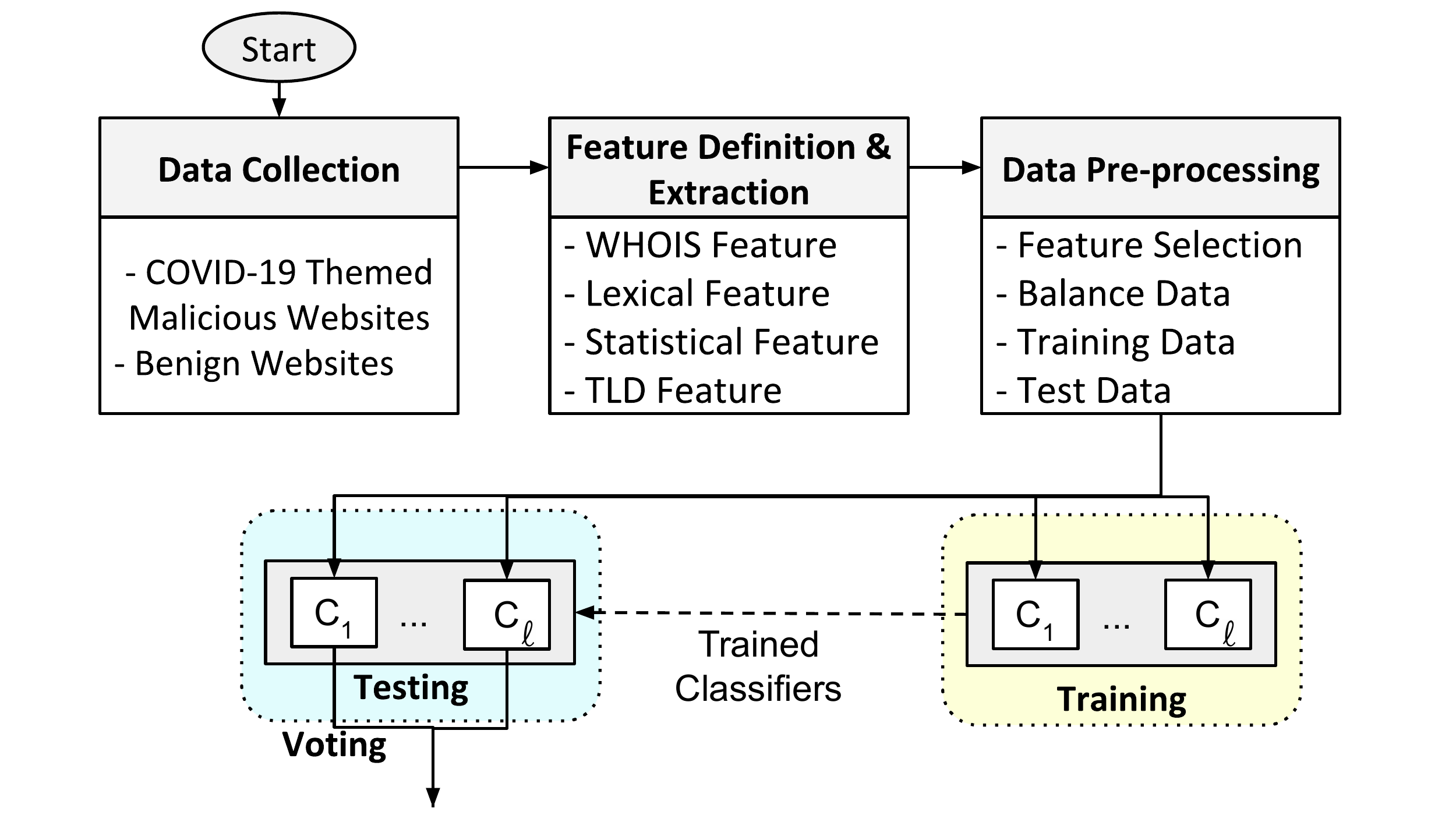}
\vspace{-2em}
\caption{Methodology for detecting COVID-19 themed malicious websites} 
\label{fig:detectionFramework}
\end{figure}

We propose leveraging machine learning to detect COVID-19 themed malicious websites and answer:
\begin{itemize}
\item RQ5: Which classifier is competent in detecting COVID-19 themed malicious websites?
\item RQ6: What is the impact of WHOIS features on the classifier's effectiveness?
\end{itemize}
In order to answer these questions, we need to train detectors. Figure \ref{fig:detectionFramework} highlights the methodology for detecting COVID-19 themed malicious websites. The methodology can be decomposed into the following modules: data collection, feature definition and extraction, data pre-processing, classifier training, and classifier test.

Data about websites need to be collected from reliable sources. The collected data may need enrichment to provide more information, as what will be illustrated in our case study. Then, features may be defined to describe these websites. In the case of using deep learning (which requires much larger datasets), features may be automatically learned. One may consider a range of classifiers, which are generically called $C_i$'s in Figure \ref{fig:detectionFramework}. 
As shown in Figure \ref{fig:detectionFramework}, 
one can use classifiers individually or an ensemble of them (e.g., via a desired voting scheme, such as weighted vs. unweighted majority voting). In the simple form of unweighted majority voting, a website is classified as malicious if majority of the classifiers predict it as malicious; otherwise, it is classified as benign.

In order to evaluate the effectiveness of the trained classifiers, we propose adopting the standard metrics, including: accuracy (ACC), false-positive rates (FPR), false-negative rates (FNR), and $F1$-score. 
Specifically, let $TP$ be the number of true positives, $TN$ be the number of true negatives, $FP$ be the number of false positives, and $FN$ be the number of false negatives. Then, we have
ACC $= \frac{TP + TN}{TP + TN + FP + FN}$, 
FPR $= \frac{FP}{FP + TN}$,
FNR $= \frac{FN}{FN + TP}$, and
%$DR = \frac{TN}{FP + TN}$
$F1$-score $= \frac{2TP}{2TP + FP + FN}$.
%Each metric is calculated by taking the average after running experiments for 5 times with shuffled test and training data. 

\section{Case Study}
\label{sec:experiments}

Our case study applies the methodology to specific datasets. 
%Since the same dataset is used for both the characterization and detection studies, we start with its description.

\subsection{Data Collection}
Our dataset of COVID-19 malicious website examples are obtained from what was published between 2/1/2020 and 5/15/2020 by two sources:
(i) CheckPhish \cite{COVID19_data_checkphish}, which contains 131,761 malicious websites waging scamming attacks related to COVID-19; and
(ii) DomainTools \cite{COVID19_data_DT}, which contains 157,579 malicious websites waging malware, phishing, and spamming attacks related to COVID-19.
The union of these two sets leads to a total of 221,921 malicious websites, denoted by $D_{malicious}$, owing to the fact that 67,419 websites belong to both sets.
For obtaining benign websites, we use the top 250,000 websites from Cisco's Umbrella 1 million websites dataset \cite{cisco_umbrella_dataset} on 05/16/2020, denoted by $D_{benign}$, which is a source of reputable websites. We compile a merged dataset denoted by $D_{initial} = D_{malicious} \cup D_{benign}$.

In order to collect WHOIS information of a website, we use the python library {\tt whois 0.9.7} to query the WHOIS database on 8/7/2020. We observe that 42,540 (or 19.17\%) out of the 221,921 malicious websites have no WHOIS information available, and 93,082 (or 37.2\%) out of the 250,000 benign websites have no WHOIS information available. This means that the presence/absence of WHOIS information does not indicate that a website is malicious or not.

\subsection{Characterization Case Study}

\subsubsection{Answering RQ1: Identifying the WHOIS registrars that are most abused to launch COVID-19 themed malicious websites}
For this purpose, we use a subset of $D_{malicious}$ set, denoted by $D'_{malicious}$, which contains 171,901 %COVID-19 themed 
malicious websites with WHOIS {\em registrar\_name} information available.

\begin{figure}[!t]
\centering
\includegraphics[width=.91\linewidth]{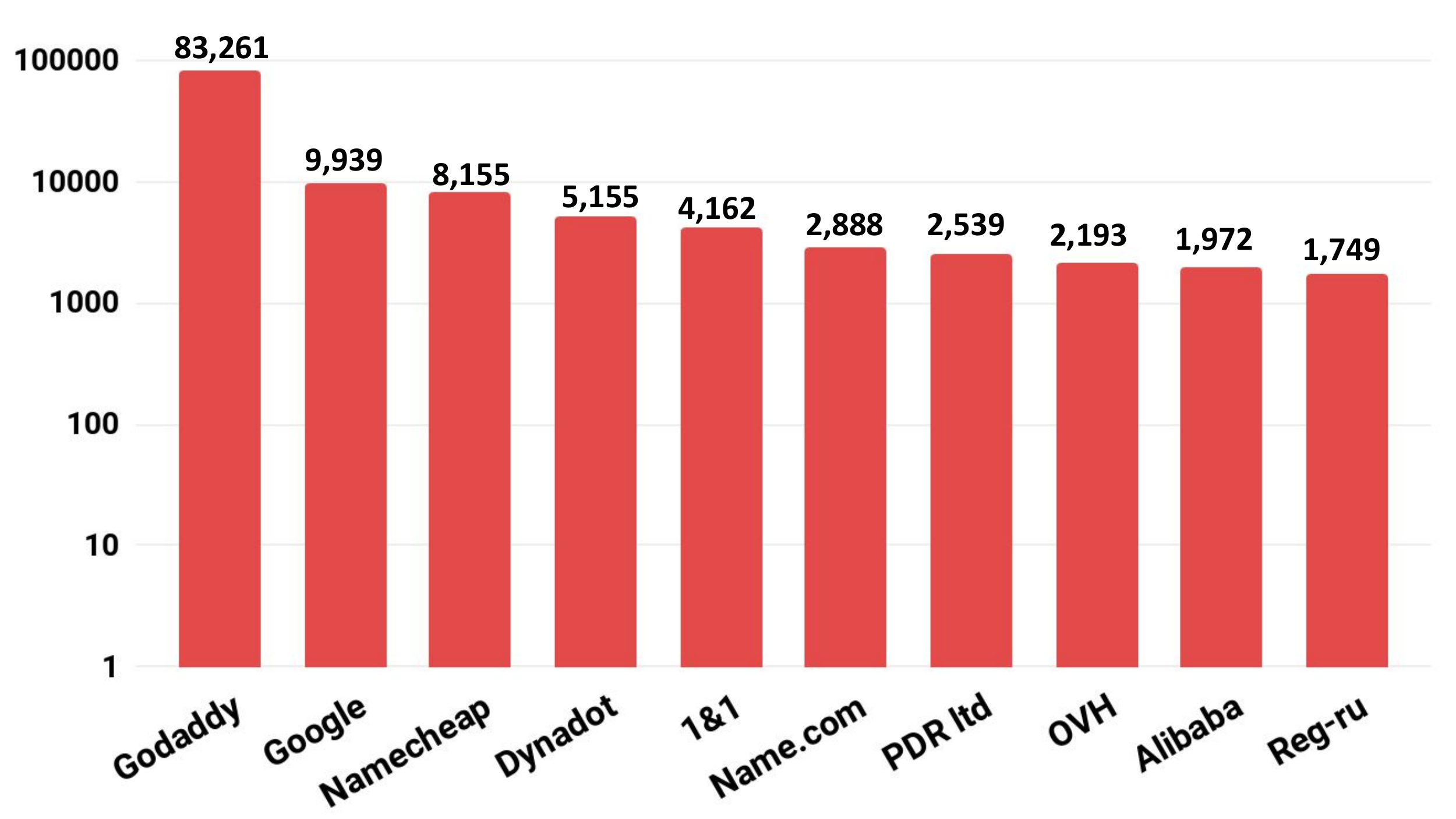}
\caption{Top 10 abused WHOIS registrars of COVID-19 themed malicious websites (the $y$-axis is in the log-scale).}
\label{fig:registrarDistribution}
\end{figure}

Figure \ref{fig:registrarDistribution} depicts the top 10 abused registrars, which are ranked according to the absolute number of 
COVID-19 themed websites in $D'_{malicious}$ that are respectively registered by them.
We observe that {\tt Godaddy} is the most frequently abused registrar, followed by {\tt Google} and {\tt Namecheap}. This finding inspires us to analyze if there is any financial incentive behind the use of a specific registrar. The cost
%and collect yearly domain registration costing for a 
registering a {\tt .com} domain in the first year, is: 
{\tt Godaddy} for \$11.99, {\tt Google} for \$9, {\tt Namecheap} for \$8.88, {\tt Dynadot} for \$8.99, {\tt 1\&1} for \$1, {\tt name.com} for  \$8.99, {\tt PDR Ltd} for \$35, {\tt OVH} for \$8.28, {\tt Alibaba} for \$7.99, {\tt Reg-ru} for \$28. This suggests that some attackers might have considered registrar {\tt 1\&1} because it is the cheapest, while some attackers use reputed registrars.

\begin{insight}
Some attackers may be incentivized to use cheaper registrars but some of the other don't.
\end{insight}

\subsubsection{Answering RQ2: Which Top Level Domains (TLDs) are most abused by COVID-19 themed malicious websites?}
In order to answer this question, we use the original dataset $D_{malicious}$, which contains 221,921 COVID-19 themed malicious websites with corresponding TLD information. % and 250,000 benign websites.
%Cisco websites) have available TLD information. 

\begin{figure}[!htbp]
\vspace{-1.5em}
\centering
\includegraphics[width=.89\linewidth]{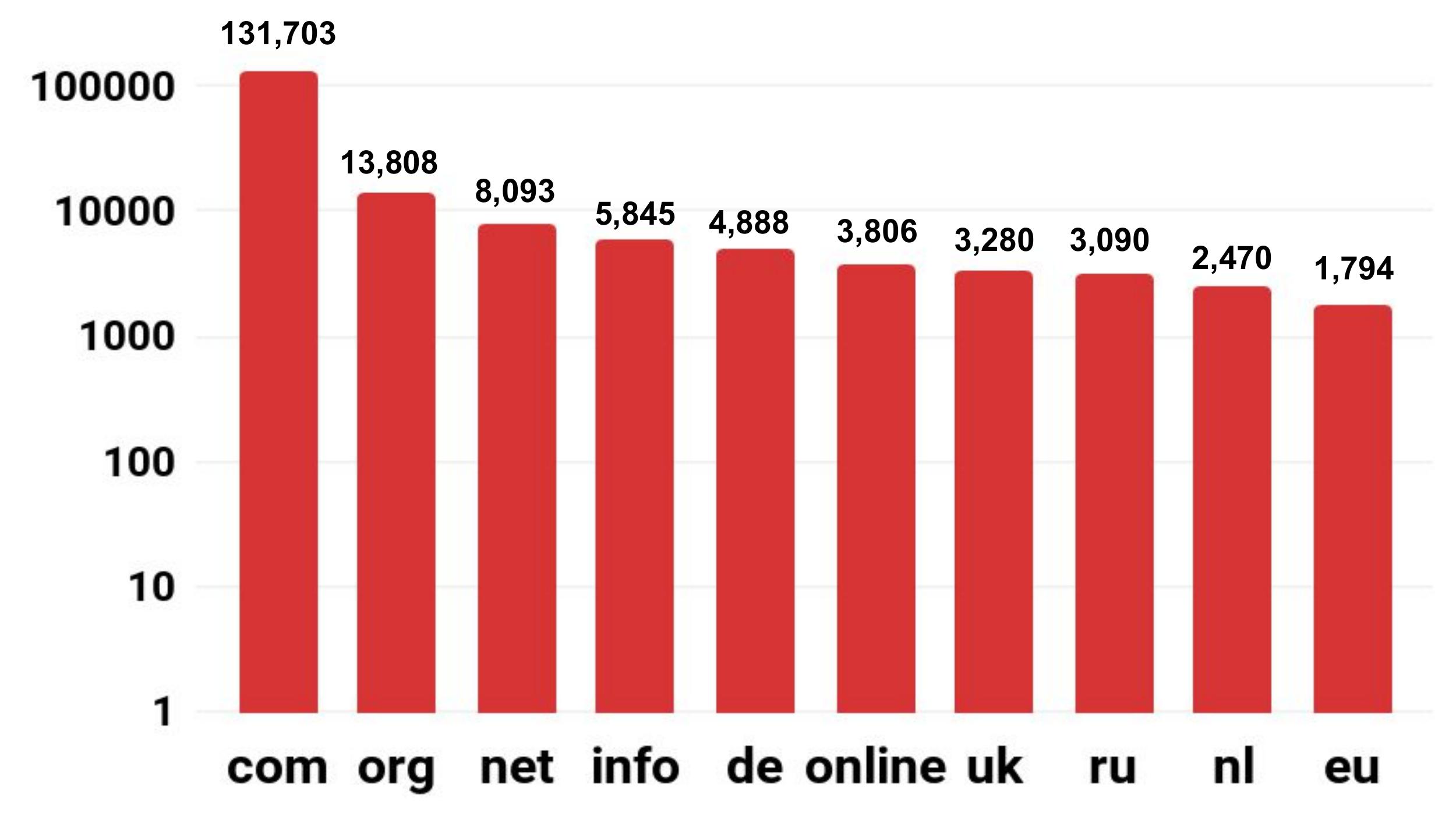}
\vspace{-1.5em}
\caption{Top 10 abused TLDs of COVID-19 themed malicious websites (the $y$-axis is in the log-scale).}
\label{fig:tldDistribution}
\end{figure}

Figure \ref{fig:tldDistribution} depicts the 
top 10 abused TLDs, which are ranked according to the absolute number TLDs for COVID-19 themed malicious websites. We make the following observations. First, {\tt .com} hosts the highest number of malicious websites, followed by {\tt .org} and {\tt .net}.
Second, 5 of the top 10 abused TLDs correspond to country-level {\tt ccTLDs}, including {\tt .de}, {\tt .uk}, {\tt .ru}, {\tt .nl} and {\tt .eu}. 
\begin{insight}
Attackers often abuse popular TLDs.
%to evade detection.
\end{insight}

\subsubsection{Answering RQ3: What trends are exhibited by COVID-19 themed malicious websites?}
\begin{figure}[!htbp]
\vspace{-1.5em}
\centering
\includegraphics[width=0.91\linewidth]{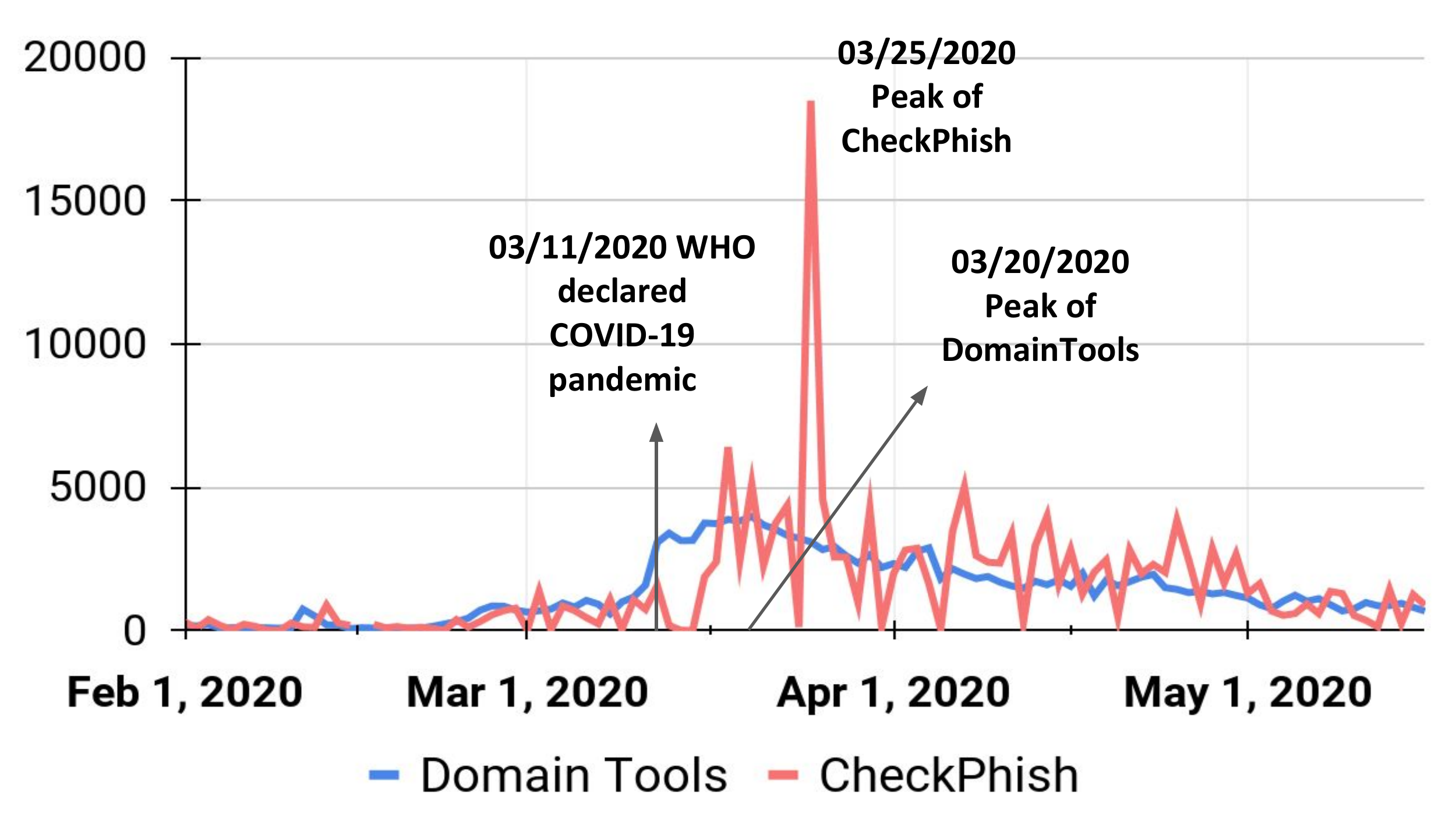}
\caption{Trends of COVID-19 themed malicious website.}
\label{fig:website_trend}
\end{figure}

In order to answer this question, we use the dataset $D_{malicious}$ mentioned above.
%, namely the COVID-19 themed malicious websites obtained from DomainTools and CheckPhish datasets from 02/1/2020 to 05/16/2020. 
Figure \ref{fig:website_trend} depicts the trend of malicious websites, leading to two observations. First, there is a discrepancy between the daily numbers of websites that are reported by the two sources. According to CheckPhish, the number of COVID-19 themed malicious websites reaches the peak on 03/25/2020, with 18,495 malicious websites; according to DomainTools, the number of COVID-19 themed malicious websites reaches a peak on 03/20/2020, with 3,981 malicious websites. This data indicates that there are reporting inconsistencies among sources and many COVID-19 themed malicious websites are created at the early stage of the pandemic when {\em uncertainties} are maximum. Second, the number of COVID-19 themed malicious websites, by and large, has been decreasing since the last week of March 2020 (i.e., two weeks after the pandemic declaration), leading to about 1,000 websites per day during the first week of May 2020 (i.e., about two months after pandemic declaration). However, there is still oscillation. One possible cause is that the attackers have been waiting to create new COVID-19 themed malicious websites based on the pandemic's new developments (e.g., vaccine). 

\begin{insight}
Inconsistencies in reporting mechanisms, attackers are agile in creating COVID-19 themed malicious websites.
\end{insight}

\subsubsection{Answering RQ4: Which theme keywords are mostly abused by attackers, and how?}

In order to answer this question, we analyze the dataset $D_{malicious}$ mentioned above.
We use the python library {\tt wordninja} with English Wikipedia language model \cite{python_wordninja} to split domain name strings and extract COVID-19 themed keywords. 
We observe that 4 keywords (i.e., {\em covid}, {\em corona}, {\em covid19}, and {\em coronavirus}) are most widely used as expected; they are followed by {\em mask}, {\em quarantine}, {\em virus}, {\em test}, {\em facemask}, {\em pandemic}, and {\em vaccine}. We  extract more than 19,000 keywords. A further analysis of the domain names reveals that attackers create COVID-19 themed malicious websites with names containing geographical attributes. For example, {\tt coronaviruspreventionsanantonio.com}, { \tt coronavirusprecentionhouston.com}, and {\tt coronaviruspreventiondallas.com} use a combination of city name and a COVID-19 themed keyword. Moreover, we observe the existence of COVID-19 themed ``parking'' websites, which have no content at the present time but might be used for upcoming COVID-19 themes.

\begin{insight}
Attackers are crafty in using COVID-19 themed keywords and geographical information in creating COVID-19 themed malicious website domain names.
\end{insight}

\subsection{Detection Case Study}

Given $D_{initial}$, the detection case study proceeds as follows.

\subsubsection{Feature Definition and Extraction}
We define features according to the following aspects of websites: WHOIS (F1-F4), domain name lexical information (F5-F9), statistical information (F10), and Top-Level Domain or TLD (F11).
\begin{itemize}
\item Current WHOIS registration lifetime (F1): This is the number of days that has passed since a website's registration, with respect to the date when this feature's value is extracted  (e.g., 08/07/2020 in our case).

\item Remaining WHOIS expiration 
lifetime (F2): This is the number of remaining days before a website's WHOIS registration expires, with respect to the date when this feature's value is extracted (e.g., 08/07/2020 in our case).  

\item Number of days since last WHOIS update (F3): This is the number of days elapsed since a website's last update with respect to the date when this feature's value is extracted (e.g., 08/07/2020 in our case).

\item WHOIS registrar reputation (F4): We propose measuring a WHOIS registrar's reputation as $\frac{n}{|D_{benign}|}$, where $n$ is the number of benign websites in $D_{benign}$ that are registered by this particular registrar and $|D_{benign}|$ is the size of set $D_{benign}$. 

\item Number of dots in domain name (F5): This is the number of dots (character `.') in the domain name. For example, domain {\tt any.com} has 1 dot.

\item Domain hyphen count (F6): This is the number of hyphens (`-') in a domain name. 

\item Domain vowel count (F7): This is the number of vowels (i.e., {\em a}, {\em e}, {\em i}, {\em o}, {\em u}) in a domain name. 

\item Domain digits percentage (F8): This is the ratio of the number of digits (0-9) in a domain name to the number of characters including digits.

\item Domain unique alphabetic-numeric characters count (F9): This is the total number of unique alphabetic and numeric characters (i.e., a-z, A-Z, 0-9) in a domain name.

\item Domain entropy (F10): This is the Shannon entropy \cite{shannon_entropy} of the domain name (i.e., a kind of statistical information), which is computed based on the frequency of characters in the domain name.

\item TLD Reputation (F11): We propose measuring a TLD's reputation as 
$\frac{m}{|D_{benign}|}$, 
where $m$ is the number of websites in $D_{benign}$ that contain this particular TLD.
\end{itemize}

\subsubsection{Data Pre-Processing}
Given that some websites may not have information for the features, it is important to consider different scenarios. In our example, we propose considering two datasets that can be derived from  $D_{initial}$ because some websites do not have information for the WHOIS features.
\begin{itemize}
\item Dataset $D_1 \subset D_{initial}$ consists of websites for which WHOIS information is available (i.e., %values of 
features F1-F4 are available). $D_1$ contains 21,749 websites in total, including 16,411 COVID-19 themed malicious websites and 5,338 benign websites.

\item Dataset $D_2 \subset D_{initial}$, where $D_1 \cap D_2 = \emptyset$, consists of websites for which WHOIS information is absent (i.e., %values of 
features F1-F4 are entirely missing). $D_2$ contains 135,621 websites, including 42,540 malicious websites and 93,081 benign websites. For each website belonging to $D_2$, only values of the 7 features (i.e., F5-F11) are available.

%{\color{red}%\item Dataset $D_3$ consists of websites for which WHOIS feature F4 is available but the other WHOIS features (i.e., F1-F3) may or may not be available, meaning that $D_1 \subseteq D_3 \subseteq D_{initial}$ and $D_2 \cap D_3 = \emptyset$. It turns out $D_3$ contains 179,381 COVID-19 themed malicious websites and 156,918 benign websites.} 
%\footnote{{\color{orange}do not need $D_3$ anymore as it was only used in Answering RQ1, can remove this, I have edited description in RQ1...?}}%{\color{red}This dataset is effectively used for incorporating each WHOIS registrars' {\em badness}.}
\end{itemize}

%Given these datasets, we investigate feature selection and address data imbalance as follows. 
%Before answering RQ5 and RQ6 as a data pre-processing step 

\begin{table} [!htbp]
\caption{Relative importance of features in $D_1$ with respect to the random forest method.}
  \label{tab:featureSelection}
  \begin{center}
      \begin{tabular}{|c|c||c|c|}
        \hline
        Feature & Importance & Feature & Importance \\
        \hline
        F1 &  0.429 & F7  & 0.080 \\
        \hline
        F2 & 0.094 & F8 & 0.009\\
       \hline
        F3 & 0.131 &  F9 & 0.028  \\
       \hline
        F4 & 0.065 & F10 & 0.029 \\
       \hline
        F5 & 0.065 & F11 & 0.068 \\
       \hline
        F6 & 0.003 &  &\\
       \hline
      \end{tabular}
  \end{center}
\end{table}

Since only $D_1$ contains all WHOIS information, We use it for feature selection study. For this purpose, we use the {\em random forest classification feature importance} method \cite{kuhn2013applied} (with the 80-20 splitting of training-test data) to 
find the important features. 
Table \ref{tab:featureSelection} depicts the relative importance of the features in $D_1$. We observe that F6 and F8 have a very small relative importance (i.e., $< 0.01$) when compared to the others, suggesting that hyphens and digits are equally used in malicious or benign domain names.
Hence, we will eliminate F6 and F8 in the rest study of $D_1$. 

In order to see whether or not the feature selection result is impacted by the data imbalance of $D_1$ (with the malicious:benign ratio being 3.1:1), we explore two widely-used methods: (i) {\em oversampling} the minority class to replicate some random examples; and (ii) {\em undersampling} the majority class to remove some random examples. At first, we do the 80-20 splitting of training-test  data, and then change the malicious:benign ratio in the training set, while keeping the test set intact. %We want to see how a classifier' effectiveness changes with the training ratio. 
We wish to identify the ratio that achieves the highest $F1$-score.
In what follows we only report the results of Random Forest because it outperforms the other classifiers for the original dataset $D_1$.

\begin{table}[!htbp]
\caption{Impact of the malicious:benign ratio on the effectiveness of the Random Forest classifier with {\em Oversampling} and {\em Undersampling}, where $D_1$ with ratio 3.1:1 is the original $D_1$.}
  \label{tab:RatioParameter}
\vspace{-2em}
  \begin{center}
      \begin{tabular}{|c|c|c|c|c|c|c|c|c}
        \hline
        Dataset & Method & {\em Ratio} & ACC & FPR & FNR & $F1$-score \\
        \hline
        $D_1$ & (none) & 3.1:1 & 0.980 & 0.030 & 0.017  & 0.987\\
        \hline
        $D_1$  & Oversample & 2:1& 0.980 & 0.030 & 0.018  & 0.986  \\
        \hline
        $D_1$  & Oversample & 1.67:1 & 0.980 & 0.027 & 0.017 & 0.988 \\
        \hline
        $D_1$ & Oversample & 1.43:1 & 0.979 & 0.028 & 0.019  & 0.986\\
        \hline
        $D_1$ & Oversample & 1.25:1 &  0.979 & 0.028 & 0.018 & 0.986\\
        \hline
        $D_1$ & Oversample & 1.11:1 & 0.979 & 0.027 & 0.019  & 0.986\\
        \hline
        $D_1$ & Oversample & 1:1 & 0.979 & 0.026 & 0.019  & 0.986\\
        
        \hline \hline 
        $D_1$  & Undersample & 2:1 & 0.977 & 0.023 & 0.022  & 0.985 \\
        \hline
        $D_1$  & Undersample & 1.67:1 & 0.976 & 0.023 & 0.025 & 0.984 \\
        \hline
        $D_1$ & Undersample & 1.43:1 & 0.975 & 0.023 & 0.025 & 0.984 \\
        \hline
        $D_1$ & Undersample & 1.25:1 & 0.972 & 0.020 & 0.031 & 0.981\\
        \hline %\hline
     %   $D_1$ & 0.8 & 0.935 & 0.944 \\
     %   \midrule
%        $D_2$ & - & 1:2.2 & 0.944 & 0.047 & 0.074  & 0.913  \\
   %     \hline
        %$D_2$ & 0.5 & 0.946 & 0.920 \\
        %\midrule
      %  $D_2$ & 0.6 & 0.945 & 0.928 \\
      %  \midrule
%        $D_2$ & 0.7 & 0.946 & & & 0.935 \\
 %       \hline
 %       $D_2$ & Oversample & 1:1.25 & 0.944 & 0.061 & 0.050 & 0.938 \\
        %\hline
  %      $D_2$ & Oversample & 1:1.11 & 0.946 & 0.059 & 0.049 & 0.944 \\
        %\hline
   %     $D_2$ & Oversample & 1:1 & 0.947 & 0.061 & 0.043 & 0.948 \\
     %   \hline
      %  \hline
    %    $D_2$ & Undersample & 1:1.25 & 0.943 & 0.062 & 0.052 & 0.936   \\
       % \hline
        %$D_2$ & Undersample & 1:1.11 & 0.943 & 0.063 &  0.051 & 0.940 \\
        %\hline
        %$D_2$ & Undersample & 1:1 & 0.947 & 0.069 &  0.038 & 0.948 \\
        %\hline
      \end{tabular}
  \end{center}
\end{table}

Table \ref{tab:RatioParameter} shows the impacts of the malicious:benign ratio in the training set. We observe that the oversampling-incurred ratio 1.67:1 leads to the highest $F1$-score (and the second best FPR and lowest FNR), while undersampling never performs better than the original data ratio in terms of accuracy and $F1$-score. This can be explained by the fact that the latter eliminates useful information.
This prompts us to use oversampling to achieve the 1.67:1 ratio when training classifiers, which turns $D_1$ into $D'_1$ (i.e., the training set is augmented).

\begin{figure}[!htbp]
\centering
\includegraphics[width=.33\textwidth]{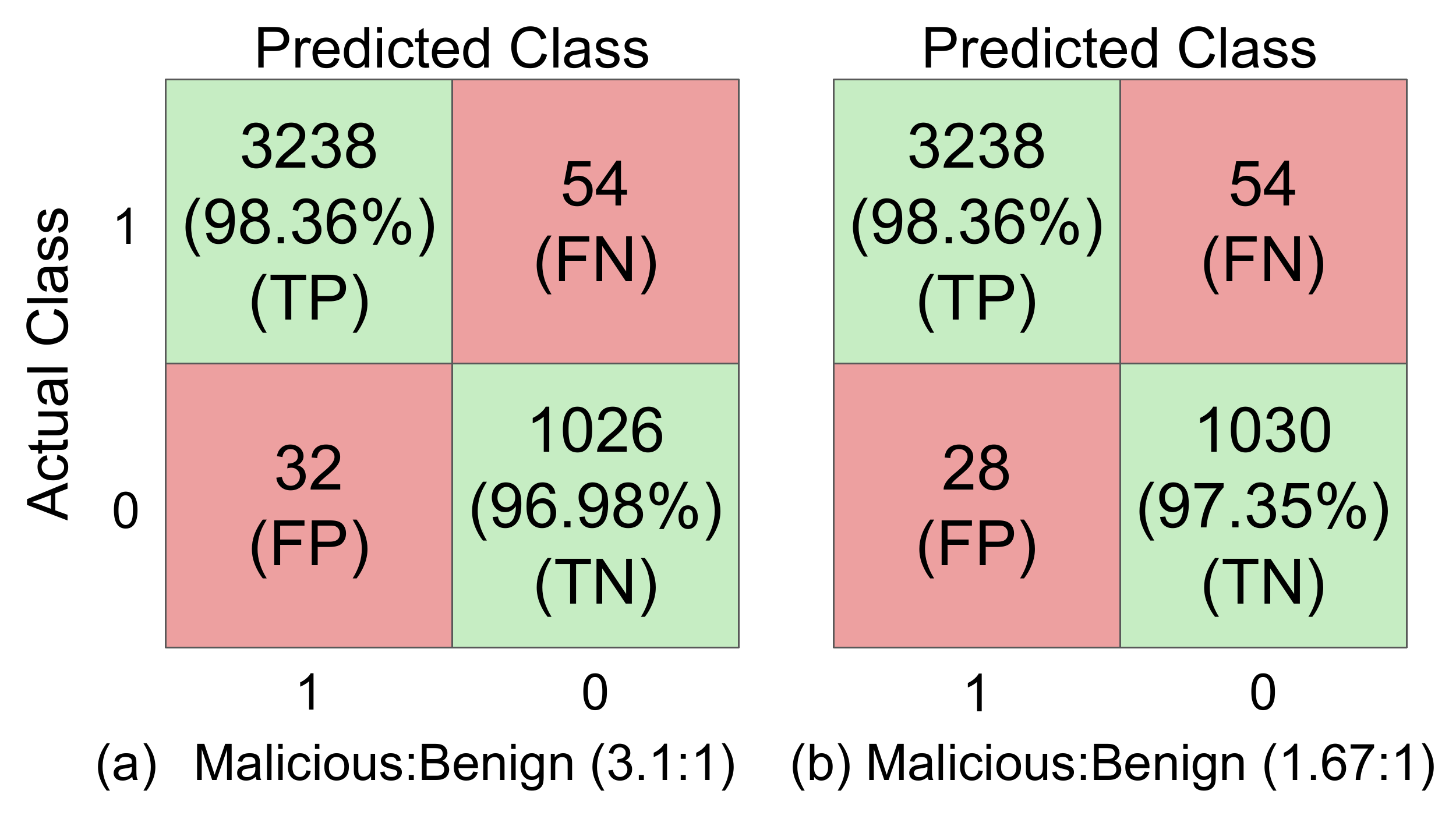}
\caption{Confusion matrix for (a) $D_1$ with 3.1:1 malicious:benign ratio in the training data and (b) $D'_1$ with 1.67:1 ratio in the training data.}
\label{fig:confusionMatrix}
\end{figure}

Figure \ref{fig:confusionMatrix} further highlights the {\em confusion matrix} of the experiment one the same test set but corresponding to $D_1$ and $D'_1$, which shows a slight improvement in detection when augmenting the training set with oversampling. 
%Since the accuracy and $F1$-score is similar for both $D_1$ and $D'_1$, the feature importance have not been affected by the class imbalance. 

\begin{insight}
The data imbalance issue does not affect the model performance significantly
in this case, perhaps because the degree of imbalance is not severe enough.
\end{insight}

\subsubsection{Training and Test}

Having addressed the issue of feature selection and data imbalance, %and giving that our dataset is small, 
we consider the following classifiers: Random Forest (RF), Decision Tree (DT), Logistic Regression (LR), $K$-Nearest Neighbor (KNN), and Support Vector Machine (SVM). Specifically, we use the python {\tt sklearn} module to import the following classifier algorithms: (i) Random Forest or RF with parameters {\tt n\_estimator}=$100$ (i.e., 100 trees in a forest) and {\tt criterion}={\em `entropy'} (i.e., entropy is used to measure information gain); (ii) $K$-Nearest Neighbor or KNN, with parameters {\tt n\_neighbors}=$8$ (i.e., 8 of neighbors are considered), {\tt metric}={\em `minkowski'} with $p=2$ (i.e., the Minkowski metric with $p=2$ measures the distance between two feature vectors), and the rest parameters are the default values; (iii) Decision Tree or DT with default parameters; (iv) Logistic Regression or LR with default parameters; (v) Support Vector Machine or SVM with {\tt linear} kernel and other default parameters. For voting the outputs of the five classifiers mentioned above, we use the {\tt VotingClassifier()} function and set {\tt voting}={\em `hard'} (i.e., majority voting).
We always considering the 80-20 splitting of the scaled training-test data. 

%split (i.e., 80\% data are randomly chosen for training and the rest 20\% for test). 

\subsubsection{Answering RQ5 and RQ6}

In order to answer RQ5 and RQ6, we conduct the following experiments, where we use the 80-20 train-test splitting of $D_1$ and then augmenting the training set as mentioned above.
Our experiments are conducted on a virtual machine on \url{https://www.chameleoncloud.org/}, running CentOS 7 on a machine of an  x86\_64 processor with 48 cores and CPU frequency 3.1 GHz.

\begin{itemize}
    \item Experiment (Exp.) 1: Use the lexical, statistical, and TLD features (i.e., F5, F7, F9-F11) only, while ignoring the WHOIS features. (This experiment is equally applicable to $D_2$, which is not reported owing to space limitation.)
    \item Experiment (Exp.) 2: Use the WHOIS features (i.e., F1-F4), while ignoring all other features.
    \item Experiment (Exp.) 3: Use both lexical and WHOIS features (i.e., F1-F5, F7, F9-F11).
\end{itemize}

\begin{table} [!htbp]
\caption{Experimental results on dataset $D'_1$ with a range of classifiers (with oversampling), their total CPU times for training and test: Exp. 1 uses lexical features only; Exp.2 uses WHOIS features only; Exp. 3 uses both lexical and WHOIS features.  \label{tab:MLClassifierPerformance}}
\vspace{-2em}
  \begin{center}
      \begin{tabular}{|c|c|c|c|c|c|c|}
        \hline
        Exp. & Classifier & ACC & FPR & FNR & $F1$-score & Execution  \\
        & & & & & & Time(s)\\ 
        \hline
        1 & RF & 0.924 & 0.150 & 0.052 & 0.950 & 0.48\\
        \hline
        2 & RF & 0.977 & 0.025 & 0.023 & 0.985 & 0.59 \\
        \hline
        3 & RF & 0.980 & 0.027 & 0.017 & 0.988 & 0.64\\
        \hline
        1 & KNN & 0.887 & 0.199 & 0.086 & 0.925 & 0.40\\
        \hline
        2 & KNN & 0.949 & 0.034 & 0.056 & 0.966 & 0.25 \\
        \hline
        3 & KNN &  0.947 & 0.031 & 0.060 & 0.964 & 0.30\\
        \hline
        1 & DT & 0.917 & 0.151 & 0.061 & 0.945 & 0.07\\
        \hline
        2 & DT & 0.973 & 0.045 & 0.022 & 0.982 & 0.08 \\
        \hline
        3 & DT & 0.974 & 0.051 & 0.019 & 0.983 & 0.14\\
        \hline
        1 & LR& 0.885 & 0.216 & 0.082 &0.924 & 20.30\\
        \hline
        2 & LR & 0.883 & 0.362  & 0.038 & 0.926 & 23.03\\
        \hline
        3 & LR & 0.918 & 0.178 & 0.051 & 0.946 & 44.40\\
        \hline
        1 & SVM & 0.888 & 0.220 & 0.078 & 0.925 & 1.69\\
        \hline
        2 & SVM & 0.881 & 0.373 & 0.038 & 0.924 & 1.68\\
        \hline
        3 & SVM & 0.920 & 0.164 & 0.054 & 0.946 & 2.38\\
        \hline
        1 & Ensemble & 0.916 & 0.171 & 0.056 & 0.945 & 21.40\\ 
        \hline
        2 & Ensemble & 0.962 & 0. 031 & 0.041 & 0.974 & 24.75\\
        \hline
        3 & Ensemble & 0.970 & 0.035 & 0.028 & 0.980 & 45.70\\
     %   \hline
%        1 & Ensemble (best 3) & 0.913 & 0.170 & 0.046 & 0.936 \\
    %    \hline
 %       2 & Ensemble (best 3) & 0.972 &  0.020 & 0.032 & 0.978 \\
   %     \hline
  %      3 & Ensemble (best 3) & 0.979 & 0.019 & 0.023 &  0.984\\
        
        \hline
      \end{tabular}
  \end{center}
\end{table}

Table \ref{tab:MLClassifierPerformance} summarizes the experimental results with a range of classifiers and the actual time spent on training a model and classifying the entire test set. 
%,  {\color{orange}%where we do 80-20 train-test split of $D_1$, and then update the training data with augmented $D'_1$ (where $D_1 \subset D'_1$), while keeping the test data unchanged. This keeps us safe from overestimating the model with duplicate entries in the test data. 
%where our training includes %80\% of the dataset are used for training (including 
%13,118 malicious and 7,870 benign websites and the %rest 20\% are used for 
%test %(
%includes 3,292 malicious and 1,058 benign websites.} 
We make several observations. 
First, for a specific classifier, using WHOIS features alone (Exp. 2) almost always leads to significantly higher effectiveness than using lexical features alone (Exp. 1), except for Logistic Regression. Second, for a fixed classifier, using both lexical and WHOIS features together (i.e., Exp. 3) always performs better than using lexical or WHOIS features alone. 
%This means that the WHOIS features do carry useful information about the websites. Second, when WHOIS information is available, using lexical features alone leads to a high FPR, which is reduced significantly with {\em RF},{ \em KNN}, and {\em DT} classifiers when WHOIS information is considered. This means WHOIS features carry more value than lexical features.
Third, among the classifiers considered, Random Forest performs the best in every metric %{\color{orange}including the CPU times} 
in each experiment. 
%This means that Random Forest {\color{orange}is faster and achieves the best detection} performance when using both lexical and WHOIS features, {\color{orange}followed by Decision Tree.}
In particular, Random Forest (i.e., non-linear classifier) achieves a better performance than the Ensemble method because there are classifiers (e.g., Logistic Regression and SVM) that are substantially less accurate than the other classifiers and therefore ``hurt'' the voting results. Fourth, Decision Tree has the fastest execution time, followed by KNN and Random Forest, while Logistic Regression is the slowest and causes a delay for the voting ensemble. 
%{\color{orange}Additionally, Exp. 1 is equally applicable to $D_2$ dataset or any augmented $D'_2$ with minority class oversampling. 
To understand the generalizability, when conducting Exp. 1 on the augmented $D'_2$ with the benign:malicious ratio at 1.25:1, we observe that Random Forest outperforms other models by achieving a 0.947 accuracy, a 0.066 FPR, a 0.041 FNR, and a 0.947 F1-score.

\ignore{
{\color{green}
Table \ref{tab:MLClassifierPerformance} summarizes the experimental results with each classifier and the ensemble voting result.
%depicts the performance of the proposed detection methodology in the context of dataset $D_1^\prime$ with both cases, (i) WHOIS features absent (step-1), (ii) WHOIS features present (step-2). 
We make several observations. 
First, for a fixed classifier, using WHOIS features (i.e., Exp. 2) almost always leads to a significantly higher effectiveness than not using the WHOIS features (i.e., Exp. 1), except for the F1-score of the Logistic Regression. This means that the WHOIS features do carry useful information about the websites. 
Second, Decision Tree achieves the highest accuracy in every metric, especially when the WHOIS features are used. In particular, Decision Tree is significantly more accurate than the Ensemble method, which can be attributed to the fact that there are classifiers (e.g., Logical Regression and SVM) that are substantially less accurate than Decision Tree and therefore ``hurt'' the voting results.}
}

\begin{insight}
COVID-19 themed malicious website detectors must consider WHOIS features; and Random Forest performs the best among the classifiers that are considered.
\end{insight}

\section{Discussion}
\label{sec:discuss}

The present study has several limitations, which should be addressed in future studies. First, we use a heuristic method to determine the ground truth. This heuristic method can only approximate the ground truth because the data sources (i.e., CheckPhish and DomainTools feeds 
%and Cisco Umbrella 
in this case) may contain some errors. 
%Our methodology can successfully detect COVID-19 themed malicious websites effectively. However, there are still challenging areas to improve, like, reducing {\em false-positive-rates}. At first, we can argue on the quality and the collection methodology (e.g., the inclusion of users' reported scam websites by CheckPhish) of the ground truth datasets (e.g., CheckPhish, DomainTools feeds) for a bit higher {\em false-positive-rates}, as these datasets are the baseline of our methodology. However, given the scarcity of the ground truth datasets, this is a challenging problem to cope with. 
Second, we could not avoid the data imbalance problem, meaning that 
%wherein real-world scenarios, it is harder to find equal size datasets for both malicious and benign websites; hence there is a chance 
the resulting detectors or classifiers may be slightly biased towards the majority class even after the oversampling. 
%Moreover, for the Benign websites, we depend on a ranked websites' list, where going too far down the list might include websites that are not to be considered reputed anymore. 
Third, we only considered the WHOIS and URL lexical features, but not the website contents or the network layer features, Fourth, we only considered five WHOIS features because most of the other kinds of WHOIS information are largely missing, which means that WHOIS registrars need to collect more detailed information than what is presented at the moment of writing. Fifth, application of deep learning models or explainable ML are left to future research. Sixth, we observe that the python library {\tt wordninja} can make bad splits at times (e.g., when a domain name is seemingly in English characters but actually in another languages).

\section{Conclusion}
\label{sec:conclusion}

We have presented the first systematic study on {\em data-driven} characterization, and detection of COVID-19 themed malicious websites. We presented a methodology and applied it to a specific dataset. Our experiments led to several insights, highlighting that attackers are {\em agile}, {\em crafty},  {\em economically incentivized} in waging  COVID-19 themed malicious websites attacks. Our experiments show that Random Forest can serve as an effective detector against these attacks, especially when WHOIS information about websites in question is available. This highlights the importance of domain registrars to collect more information when registering domains in future.

\smallskip

\noindent{\bf Acknowledgement}. We thank the reviewers for their useful comments. This work was supported in part by ARO Grant \#W911NF-17-1-0566, ARL Grant \#W911NF-17-2-0127, and the NSA OnRamp II program.

\bibliographystyle{IEEEtran}
\bibliography{sample-base,metrics}

\ignore{
\appendix
\label{sec:appendixA}
%\label{sec:appendix}
\begin{figure*}[!htbp]
\centering
\includegraphics[width=1.01\linewidth]{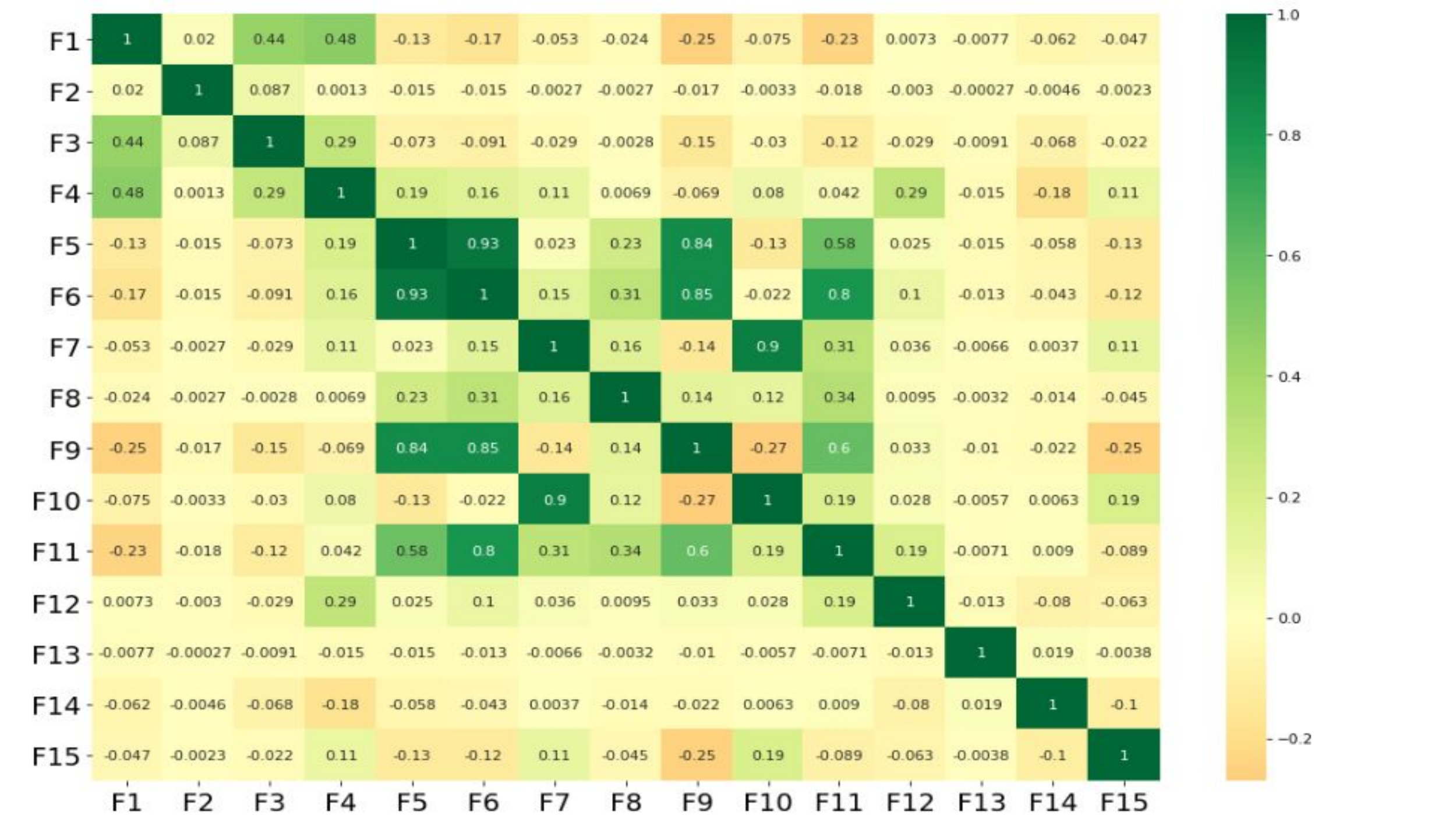}
\caption{Color-coded Pearson correlation matrix among the features in $D_1$ (green indicates strong positive correlation, yellow indicates weak correlation, red indicates strong negative correlation). 
}
\label{fig:feature_corr_D1}
\end{figure*}

\begin{figure*}[!htbp]
\centering
\includegraphics[width=1.01\linewidth]{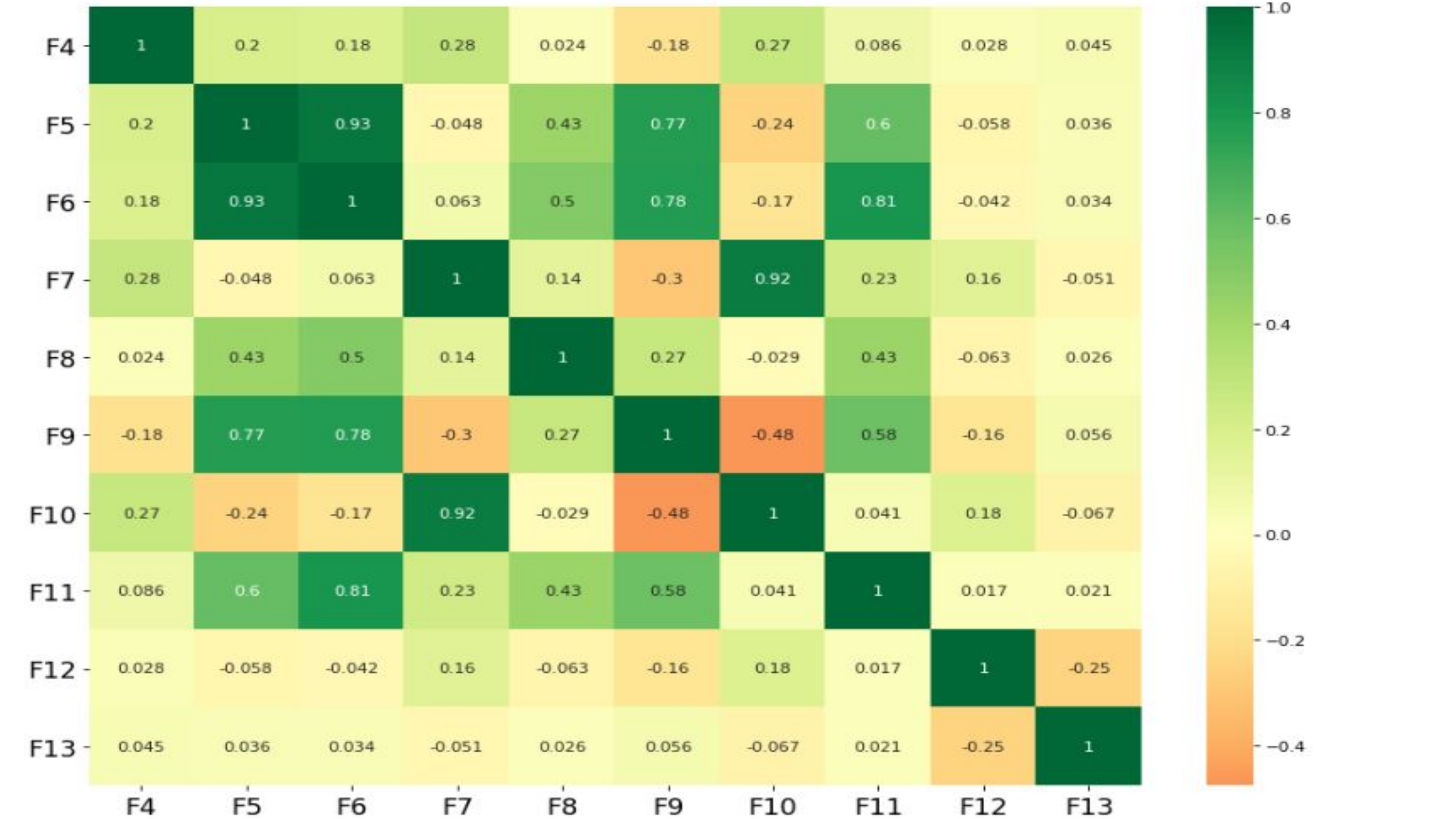}
\caption{Color-coded Pearson correlation matrix among the features in $D_2$ (green indicates strong positive correlation, yellow indicates weak correlation, red indicates strong negative correlation).}
\label{fig:feature_corr_D2}
\end{figure*}
}

\end{document}